\documentclass{elsart}
\usepackage{graphicx}

\begin{document}
\begin{frontmatter}
\title{Cosmogenic neutrinos as a probe of the transition from Galactic to extragalactic cosmic rays}

\author[utap]{Hajime Takami}\footnote{E-mail addresses: takami@utap.phys.s.u-tokyo.ac.jp (H.Takami), kmurase@yukawa.kyoto-u.ac.jp (K.Murase), nagataki@yukawa.kyoto-u.ac.jp (S.Nagataki), sato@phys.s.u-tokyo.ac.jp (K.Sato)}, 
\author[yitp]{Kohta Murase}, 
\author[yitp]{Shigehiro Nagataki}, and 
\author[utap,resceu,ipmu]{Katsuhiko Sato}

\address[utap]{Department of Physics, School of Science, the University of Tokyo, 7-3-1 Hongo, Bunkyo-ku, Tokyo 113-0033, Japan}
\address[yitp]{Yukawa Institute for Theoretical Physics, Kyoto University, Oiwake-cho, Kitashirakawa, Sakyo-ku, Kyoto, 606-8502, Japan}
\address[resceu]{Research Center for the Early Universe, School of Science, the University of Tokyo, 7-3-1 Hongo, Bunkyo-ku, Tokyo 113-0033, Japan}
\address[ipmu]{Institute of Physics and Mathematics for Universe, the University of Tokyo, Kashiwa, Chiba, 277-8582, Japan}

\begin{abstract}
There are two promising scenarios that explain the ankle, 
which is a dip in the spectrum of cosmic rays at $\sim 10^{19}$ eV. 
A scenario interprets the ankle as the transition from Galactic 
to extragalactic cosmic rays ({\it ankle-transition scenario}), 
while the other is that the dip caused by pair production 
on the cosmic microwave background radiation ({\it proton-dip scenario}). 
In this paper, we consider whether cosmogenic neutrinos can be a clue 
to judge which scenario is favored. 
We calculated the fluxes of cosmogenic neutrinos 
following these scenarios with plausible physical parameter sets, 
and found several important features as follows. 
First of all, the neutrino flux at $\sim 10^{20}$ eV becomes much higher 
in the ankle-transition scenario 
as long as the maximum energy of the cosmic rays at sources 
is sufficiently high. 
On the other hand, the neutrino spectrum has a characteristic peak 
at $\sim 10^{16}$ eV in the proton-dip scenario 
on the condition that extragalactic protons significantly contribute 
to the observed cosmic rays down to $10^{17}$ eV. 
Thus, we conclude cosmogenic neutrinos should give us a clue 
to judge which scenario is favored, 
unless these features are masked by the neutrino background 
coming from possible, powerful neutrino sources such as AGNs and GRBs. 
We also found an interesting feature 
that the neutrino flux at $\sim 10^{18}$ eV depends only 
on the cosmological evolution of the cosmic ray sources. 
That means cosmogenic neutrinos with the energy bring us information on 
the cosmological evolution of the sources of ultra-high energy cosmic rays. 
Finally, we compare the fluxes of cosmogenic neutrinos 
with the expected sensitivity curves of several neutrino detectors, 
and conclude the detection of cosmogenic neutrinos in the near future 
is promising.
\end{abstract}
\begin{keyword}
Ultra-high-energy cosmic rays; Ultra-high-energy neutrinos

\end{keyword}
\end{frontmatter}

\section{Introduction} \label{introduction}

What energy is the transition point of Galactic cosmic rays (GCRs) 
and extragalactic cosmic rays (EGCRs) in cosmic ray spectrum 
is an intriguing problem in cosmic ray astrophysics. 
The observed spectrum of cosmic rays, 
over more than 12 orders of magnitude in energy, 
can be described by a power-law shape 
with several spectral breaks \cite{stanev04}. 
The spectral breaks are interpreted as the transition points of sources 
to mainly contribute to the observed flux of cosmic rays. 
A spectral break at $\sim 10^{15.5}$ eV, so-called {\it knee}, 
has been thought of as the appearance of a maximum acceleration energy 
of protons at energetic objects in our Galaxy, like supernova remnants.

At $\sim 10^{19}$ eV, a spectral dip, called {\it ankle}, has been observed. 
Traditionally, the ankle has been interpreted as a transition point 
from GCRs with a steep spectrum ($\propto E^{-3.1}$) 
to EGCRs with a harder spectrum ($\propto E^{-2.0\sim-2.3}$). 
This is partly because cosmic rays above $10^{19}$eV could not be confined 
in the Galaxy by the Galactic magnetic field 
and their arrival distribution is highly isotropic. 
We call this traditional scenario {\it ankle-transition scenario} 
throughout this paper. 
In this scenario, it is an open problem how GCRs are accelerated 
up to $10^{19}$eV in our Galaxy. 
On the other hand, it has been suggested recently that extragalactic protons 
with a steep injection spectrum ($\propto E^{-2.6\sim-2.7}$) 
can reproduce the ankle as a spectral dip due to Bethe-Heitler pair-creation 
with cosmic microwave background (CMB) photons \cite{berezinsky05,aloisio07}.
This scenario, called {\it proton-dip scenario} throughout this paper, 
requires extragalactic protons at least down to $10^{18}$ eV. 
In the proton-dip scenario, the {\it second knee}, 
which is another spectral break at $\sim 10^{17.7}$ eV, 
is interpreted as a transition point from GCRs to EGCRs. 
It is an intriguing problem for the determination of cosmic ray origin 
which scenario is favored.

It is difficult to distinguish the two scenarios 
based on the energy spectrum of cosmic rays in observations. 
One of the key clues to determine the transition point 
is cosmic ray composition. 
In the ankle-transition scenario, the composition of GCRs at $\sim 10^{19}$eV 
is expected to be dominated by heavy nuclei 
because of their capability of accelerating up to such high energy, 
while proton-dominated composition is required in the proton-dip scenario. 
Thus, composition measurements may provide us with useful information 
on the transition scenarios. 
However, the results of composition measurements are difficult to constrain 
transition scenarios because the uncertainty of hadronic interaction models 
in the simulations of extensive air showers obstructs 
the accurate determination of the composition \cite{abbasi05b,unger07}.

In this study, we focus on another key element, cosmogenic neutrinos, 
which are generated by photopion production between 
ultra-high-energy cosmic rays (UHECRs) propagating in intergalactic space and 
cosmic background photons, 
and through the successive decay of producing pions and muons. 
Cosmogenic neutrinos are a definite probe of EGCRs. 
The estimations of the flux of cosmogenic neutrinos have been performed 
since the discovery of the CMB 
\cite{berezinsky69,stecker73,stecker79,yoshida93,engel01}. 
A recent work of Ref. \cite{seckel05} pointed out that cosmogenic neutrinos, 
coupled with UHECR results, would provide a sufficient description 
of the properties of UHECR sources.

Cosmic background radiation with the energy higher than the CMB 
(e.g., infrared (IR), optical, ultraviolet (UV)) can significantly contribute 
to the total flux of cosmogenic neutrinos because there is a large number 
of UHECRs which can generate neutrinos by interactions with such higher energy 
photons though the number of such photons is much smaller than the CMB 
\cite{ave05,stanev06,allard06}. 
Recent progress in the observations of high-redshift Universe 
allows constructing detailed models of the spectral energy distribution (SED) 
of IR to UV background (IR/UV below) radiation \cite{kneiske04,stecker06}. 
Refs. \cite{stanev06,allard06} discussed the neutrino flux 
using the SED model constructed by Ref. \cite{stecker06}. 
Ref. \cite{stanev06} considered the propagation of UHE protons 
and estimated the resultant flux of cosmogenic neutrinos. 
Based on this work, it was suggested in Ref. \cite{stanev06b} 
that cosmogenic neutrinos might be a key clue for determining 
the transition energy, but detailed discussions were not performed. 
On the other hand, the authors of Ref.\cite{allard06} estimated the fluxes of 
cosmogenic neutrinos on the assumptions of both a pure proton 
and mixed composition models.

Motivated by these studies, we discuss whether cosmogenic neutrinos can be 
a clue to judge which transition scenario is favored in detail in this study. 
We calculate the spectra of cosmogenic neutrinos following the two transition 
scenarios for this purpose for plausible physical parameter sets. 
Since the neutrino fluxes depend on several physical parameters, 
like the maximum acceleration energy of UHECRs, 
the minimum energy of EGCRs, the spectral shape of UHECRs, 
and the cosmological evolution of UHECR sources, 
we also investigate the parameter dependence of the neutrino fluxes 
and the capability of distinguishing between the two transition scenarios. 
We adopt an IR/UV background model other than a model used 
in Refs. \cite{ave05,stanev06,allard06}. 
The neutrino fluxes are normalized by the fluxes and spectral shapes 
of accompanying UHECRs obtained in observations. 
For simplicity, extragalactic magnetic field is neglected.

The composition of EGCRs are assumed to be purely protons in this study. 
EGCR composition is an interesting problem in itself. 
The High Resolution Fly's Eye (HiRes) reported that the composition 
of cosmic rays above $10^{18}$eV is dominated by protons as a result 
of $\left< X_{\rm max} \right>$, the averaged depth of the shower maximum, 
measurement \cite{abbasi05b}. 
A recent result by Pierre Auger Observatory (PAO) 
is consistent with the HiRes result 
within systematic uncertainty \cite{unger07}. 
On the other hand, 
studies of muon content in the extensive air shower, 
another observable for UHECR composition, 
indicate a significant fraction of heavy nuclei above $10^{19}$eV 
\cite{engel07,glushkov07}. 
The accurate interpretation of these composition measurements is difficult 
because of our poor knowledge of hadronic interactions at ultra-high-energy, 
as mentioned above. 
The PAO also reports the positional correlation 
between the arrival directions of the highest energy cosmic rays above 
$5.7 \times 10^{19}$eV and nearby active galactic nuclei (AGNs) 
\cite{pao07a,pao07c}. 
If these AGNs are really the sources of the observed events, 
this fact implies the highest energy cosmic rays are dominated by protons 
because of small deflections by the Galactic magnetic field. 
This implication is independent of the composition measurements. 
Thus, an assumption that EGCRs are dominated by protons 
up to the highest energies is reasonable.

This paper is organized as follows: 
in Section \ref{method}, we explain our calculation method 
of cosmogenic neutrino fluxes in detail. 
In Section \ref{results}, 
we address our results and discuss the detectability of cosmogenic neutrinos 
taking neutrino oscillation into account. 
In Section \ref{discussion}, 
several uncertainties on the neutrino flux are discussed. 
We conclude in Section \ref{conclusion}.

\section{Our Calculation Method} \label{method}
In this section, 
our calculation method of the flux of cosmogenic neutrinos is explained. 
The neutrino flux can be obtained by calculating the number of neutrinos 
produced by propagating protons injected from a source, 
and then by integrating such neutrinos over all sources. 
We explain the propagation of UHE protons and their interactions 
with cosmic background photons in Section \ref{method_prop}. 
In Section \ref{method_neutrinos}, our treatment of neutrino production 
is described. 
Then, we represent UHECR source models 
to calculate the total neutrino flux in Section \ref{method_integration}.

\subsection{Propagation of UHE Protons} \label{method_prop}

Protons propagating in intergalactic space interact 
with cosmic background photons 
and lose their energies through particle productions 
\cite{yoshida93,berezinsky88}. 
They also lose their energies adiabatically due to the cosmic expansion 
since they propagate over cosmological distance. 
We consider two interactions with cosmic background photons: 
photopion production and Bethe-Heitler pair creation, 
and adiabatic energy-loss.

For cosmic background photons, we take into account not only the CMB photons 
but also IR/UV background photons. 
Background photons with the energies higher than the CMB photons 
allow protons with the energies lower than the threshold of photopion 
production with the CMB to generate neutrinos. 
Thus, they increase the neutrino flux significantly 
while they do not change UHECR spectrum 
because of the smaller number density of IR/UV photons compared 
with the CMB \cite{stanev06}. 
We adopt the best-fit model of the SED of IR/UV background photons 
calculated in Ref.\cite{kneiske04}.

Photopion production is the most essential process in this study. 
This process is treated stochastically. 
The mean free path of photopion production for a proton with energy $E$, 
in an isotropic photon field, $\lambda (E,z)$, 
is calculated as \cite{protheroe96}
\begin{equation}
\frac{1}{\lambda(E,z)} = 
\frac{1}{8 \beta E^2} \int_{\epsilon_{\rm{th}}}^{\infty} d\epsilon 
\frac{1}{\epsilon^2} \frac{d n_{\gamma}}{d\epsilon}(\epsilon,z) 
\int_{s_{\rm{min}}}^{s_{\rm{max}}} ds \sigma(s) 
\left( s - {m_p}^2 c^4 \right),
\end{equation}
where $s_{\rm{min}} = ({m_{\pi^0}} + m_p)^2c^4$, 
$s_{\rm{max}} = {m_p}^2c^4 + 2 E \epsilon ( 1 + \beta)$, 
$\epsilon_{\rm{th}} = (s_{\rm{min}} - {m_p}^2c^4) 
\left[ 2E (1 + \beta)\right]^{-1}$. 
$m_p$, $m_{\pi_0}$, $c$, $\beta$, $\epsilon$, and $s$ are 
the proton mass, the neutral pion mass, the speed of light, 
the velocity of the proton in the unit of the speed of light, 
the energy of cosmic background photons, and 
the Lorentz invariant energy squared, respectively. 
$\sigma(s)$ is the total cross section of photopion production 
and $d n_{\gamma}/d\epsilon$ is the differential number density 
of cosmic background photons. 
$\sigma(s)$ is calculated by GEANT4, a Monte-Carlo simulation tool-kit 
which can simulate photomeson productions \cite{agostinelli03}. 
The GEANT4 can well reproduce the experimental total cross section 
of photopion production.

\begin{figure}[t]
\begin{center}
\includegraphics[width=0.95\linewidth]{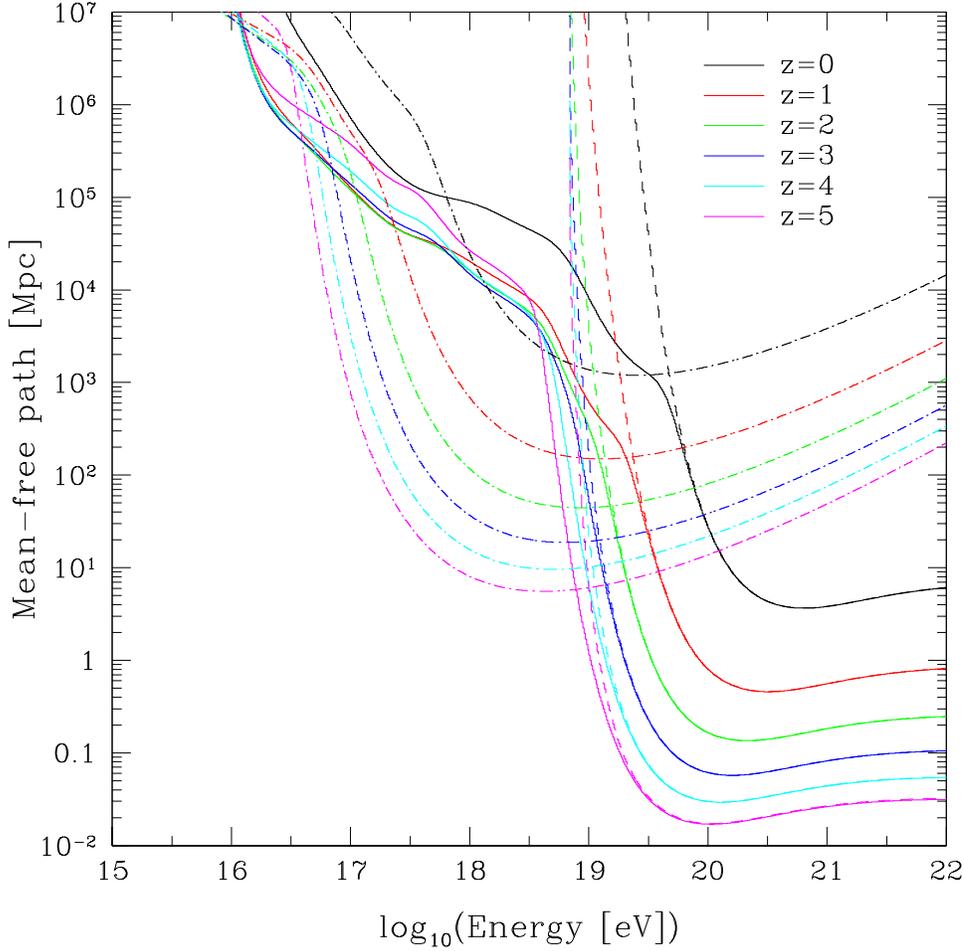}
\caption{Mean free paths of protons for photopion production 
in cosmic background radiation including IR/UV photons 
({\it solid lines}) and only CMB photons ({\it dashed lines}) at six redshifts. 
The energy-loss lengths of Bethe-Heitler pair creation of protons 
in cosmic background radiation from the CMB to IR/UV are also shown ({\it dot-dashed lines}).}
\label{fig:mfp}
\end{center}
\end{figure}

The mean free paths in intergalactic space are calculated and used 
for every $\Delta z=0.1$ up to $z=5$. 
The mean free paths at several redshifts $z$ are shown in Fig.\ref{fig:mfp}. 
The mean free path at $z=0$ rapidly decreases above $8 \times 10^{19}$eV 
where the channel of interactions with the CMB opens, 
which makes sharply cosmic ray spectrum steepening, so-called 
Greisen-Zatsepin-Kuz'min (GZK) steepening \cite{greisen66,zatsepin66}. 
At $z=0$, the minimum length of the mean free path $\lambda_{\rm min}(z)$ 
is about 4 Mpc. 
The interaction points of propagating protons are determined by a method used 
in Ref.\cite{stanev00}, based on $\lambda(E,z)$ and $\lambda_{\rm min}(z)$.

The inelasticity of photopion production, $K(s)$, is approximated 
by a conventional method 
as $K(s) = \left[ 1 - ({m_{\rm CR}}^2 - {m_{\pi}}^2)/s \right]/2$ 
for single pion production, 
where $m_{\rm CR}$ and $m_{\pi}$ are the mass of the cosmic ray 
after an interaction (proton or neutron), 
the pion mass generated in the reaction, respectively. 
For multipion production (see Section \ref{method_neutrinos}), 
we approximately adopt this formula by replacing $m_{\pi}$ with 
the total mass of pions. 
This assumption does not almost affect the flux of cosmogenic neutrinos 
shown in Section \ref{results}. 
A typical inelasticity is 0.23 at $s\sim 1.6~{\rm GeV}^2$, 
which is the Lorentz invariant energy squared at the delta resonance 
in the total cross section of photopion production.

Cosmic background photons also make protons lose their energies 
through Bethe-Heitler pair creation, $p\gamma \longrightarrow pe^+e^-$. 
Fig.\ref{fig:mfp} also shows the energy-loss lengths of this process 
which are calculated following an analytical fitting function 
given by Ref.\cite{chodorowski92}. 
This energy-loss process can be treated 
in a continuous energy-loss approximation 
since its inelasticity is very small ($2m_e/m_p \sim 10^{-3}$). 
Since the energy-loss length is much shorter than the interaction length 
of photopion production for IR/UV photons at high-redshift Universe, 
neutrino production related to IR/UV photons mainly occurs 
at about less than z=1.

The energy-loss rate of protons with energy $E$ due to the cosmic expansion 
is expressed as 
\begin{equation}
\frac{d E}{dt} = -\frac{\dot{a}}{a} E 
= -H_0 \left[ \Omega_m ( 1 + z )^3 + \Omega_{\Lambda} \right]^{1/2} E, 
\end{equation}
where $\Lambda$CDM cosmology with $\Omega_m = 0.3,~\Omega_{\Lambda}=0.7$, 
and $H_0 = 71 \rm{km~s}^{-1}~\rm{Mpc}^{-1}$ is assumed. 
We also treat this process as continuous process during propagation.

The spectra of UHE protons and cosmogenic neutrinos are calculated as follows: 
first, we consider protons with $10^{16}$-$10^{22}$eV ejected from a source. 
This energy range is divided into 60 bins with $\Delta (\log_{10} E)=0.1$ 
and 2000 protons are ejected in each bin initially 
regardless of UHECR injection spectrum. 
The propagation of protons is calculated 
taking the energy-loss processes and pion production into account, 
and then the resultant spectra of UHE protons and neutrinos at the Earth 
are recorded for each initial energy bin. 
The calculation of propagation stops when the energies of propagating protons 
reach $10^{15}$eV. 
Neutrinos are assumed to lose their energies 
only due to the adiabatic energy loss. 
Then, the resultant spectra sum up over all initial energy bins 
weighted by a given injection spectrum 
to obtain the spectra of UHE protons and cosmogenic neutrinos from a source. 
Finally, the total spectra of protons and cosmogenic neutrinos are calculated 
by integrating their spectra from a source over all sources 
taking UHECR source-evolution models into account. 
The injection spectra and UHECR source-evolution models adopted in this study 
are explained in Section \ref{method_integration}.

\subsection{Neutrino Production} \label{method_neutrinos}

Cosmogenic neutrinos are produced by the decay of charged pions generated 
by photopion production of propagating protons with cosmic background photons. 
A charged pion decays into a muon and muon neutrino, 
$\pi^+ \rightarrow \mu^+\nu_{\mu}$ 
or $\pi^- \rightarrow \mu^-\bar{\nu_{\mu}}$. 
The muon decays into an electron and neutrinos, 
$\mu^- \rightarrow e^-\bar{\nu_e}{\nu_{\mu}}$ 
or $\mu^+ \rightarrow e^+{\nu_e}\bar{\nu_{\mu}}$.

There are many reaction modes for pions in photopion production, 
since protons with sufficient energies could produce more than one pions. 
The GEANT4, which is used to calculate the total cross section, 
can also calculate exclusive cross section of each reaction mode, 
but it has some problems in parametrization. 
Therefore, we combine the experimental data of exclusive cross sections 
similarly to Ref.\cite{murase06c}. 
In addition, it takes a large CPU time 
to fully simulate photopion production one by one. 
Hence, we shall take a simpler treatment to calculate neutrino generation, 
as explained below.

We consider $p\gamma \rightarrow p\pi^0,~ n\pi^+$ 
as single pion production processes, 
and $p\gamma \rightarrow p\pi^+\pi^-,~ n\pi^0\pi^+,~ p\pi^0\pi^0$ 
as double pion production processes. 
We adopt experimental results of the cross sections of 
these reactions below $s \sim 3~\rm{GeV}^2$ 
like Refs. \cite{murase06c,schadmand03,asano06}. 
These cross sections are not confirmed well by experiments 
at $s > 3~{\rm GeV}^2$. 
Therefore, we simply extrapolate the cross sections 
from $s \sim 3$ GeV$^2$ to higher energy. 
Then, we regard the total cross section minus the contributions 
of single and double pion production 
as the cross section of triple pion production, 
which generates $\pi^0$, $\pi^+$ and $\pi^-$. 
The above treatment is sufficient 
unless multi-pion production processes are significant.

Once we judge that a propagating proton raises photopion production 
in calculation, 
the counterpart photon energy $\epsilon$, 
and $s$ are determined by a method used in Ref.\cite{protheroe96}. 
Next, it is determined which reaction is realized 
based on the probabilities proportional to 
the cross section of each reaction mode at $s$. 
The total energy of the produced pions is given by $K(s)E$. 
$K(s)E$ is the energy of the pion for single pion production. 
For double or triple pion production, 
we assume the total energy, $K(s)E$, to be divided equally into all pions. 
Generated charged pions decay into muons and muon neutrinos. 
The energies of the products are calculated 
using the two body decay algorithm. 
The energies of the products of muon decay are calculated 
using the three body decay algorithm. 
The pions and muons decay immediately.

Neutrinos are also produced by the beta decay of neutrons, 
$n \rightarrow pe^-\bar{\nu_e}$, 
which result from charged pion production. 
These neutrinos can contribute to the observed neutrino flux 
if the mean free path of the beta decay, 
$\gamma c\tau = 0.92 ( E / 10^{20} \rm{eV} )~\rm{Mpc}$, 
is shorter than that of photopion production of a neutron 
which is assumed to be equal to that of a proton. 
At $z=0$, for example, 
the mean free path of the photopion production 
is comparable with that of beta decay of a neutron with $10^{20.6}$eV. 
Hence, the neutron beta decay is also an important process 
for neutrino productions. 
This process generates only anti-electron neutrinos. 
The energies of the products are calculated 
using the three body decay algorithm.

\subsection{UHECR Source Models} \label{method_integration}

The total flux of cosmogenic neutrinos is calculated 
by integrating a neutrino flux from a source over all sources. 
The cosmological evolution of the number density 
and luminosity density of UHECR sources 
strongly affects on the flux of cosmogenic neutrinos. 
In this study, we consider four source-evolution models 
in which sources are uniformly distributed 
with cosmological evolution up to $z=5$. 

The first is a uniform distribution without cosmological 
source-evolution (UNF). 
The second is a model following the luminosity density evolution 
of quasars (QSO), which is parametrized by Ref. \cite{waxman99} 
\begin{eqnarray}
f_{\rm QSO}(z) \propto \left\{ \begin{array}{cc}
( 1 + z )^{3} & ( z < 1.3 )\\
{\rm Constant} & ( 1.3 < z < 2.7 )\\
\exp \left( 1 - \frac{z}{2.7} \right) & ( 2.7 < z )
\end{array}
\right. .
\end{eqnarray}
The third follows a star formation rate (SFR) 
deduced from the reconstruction of a IR/UV background flux used in this study, 
\begin{eqnarray}
f_{\rm SFR}(z) \propto \left\{ \begin{array}{cc}
( 1 + z )^{3.5} & ( z < 1.2 )\\
( 1 + z )^{-1.2} & ( 1.2 < z )
\end{array}
\right. . 
\end{eqnarray}
The fourth, the last model is motivated by an assumption 
that the observed UHECRs come from gamma-ray bursts 
(GRBs) \cite{waxman95}. 
Although the redshift distribution of GRBs is still controversial 
and it is different between pre-Swift and Swift bursts 
(e.g., \cite{guetta07}), 
one plausible rate history is that the GRB rate 
is more enhanced at higher redshifts 
due to metalicity effect \cite{jimenez06}. 
It infers that the progenitors of GRBs would favor metal-poor stars, 
and host galaxies of GRBs would tend to have low metalicities 
\cite{stanek06,fruchter06}. 
In this paper, we shall adopt such an evolution model, 
and refer this GRB rate model 
as a "GRB-metalicity anti-correlation" model. 
More specifically, we shall assume a following rate history, 
$f_{\rm GRB}(z) \propto (1+z)^{1.4}f_{\rm SFR}(z)$, 
which is used in Ref.\cite{yuksel06} 
based on the calculation of Ref.\cite{yoon06}. 
We use this parametrization 
in our fourth source-evolution model (SFR+GRBMAC below). 
As for the star formation rate, $f_{\rm SFR}(z)$ 
in the third model is adopted. 
Our results do not depend on the local GRB rate 
which is somewhat uncertain so far, 
because we use the observed UHECR flux 
for the normalization of the flux of cosmogenic neutrinos. 
However, note that the lower local rate requires 
higher baryon loading per GRB if UHECRs come from GRBs.

As for the proton injection spectra at all sources, 
we assume a simple power-law spectrum, 
\begin{equation}
\frac{d N}{dE} \propto E^{-\alpha} \Theta \left( E_{\rm max} - E \right) 
\Theta \left( E - E_{\rm min} \right), 
\end{equation}
where $\alpha$, $E_{\rm max}$ and $E_{\rm min}$ are a spectral index, 
the maximum acceleration energy of protons 
and the minimum energy of protons at sources respectively. 
These are treated as free parameters. 
If UHECR sources have a spectral index steeper than 2, 
the total injection energy from a source could exceed the energy budget 
of possible candidates of UHECR sources. 
Thus, $E_{\rm min}$ is required as a cutoff. 
For conserving energetics, 
a broken power-law spectrum, for example, 
$\alpha=2$ at lower energy than an energy $E_0$, 
which is a break energy of the spectral index as parameter, 
was proposed \cite{berezinsky02,berezinsky06}. 
$E_{\rm min}$ in this study should be interpreted as a minimum energy 
if UHECR injection spectrum can be expressed by a single spectral index. 

Taking these source-evolution models into account, 
the total neutrino flux from all sources is calculated as 
\begin{equation}
\frac{d N}{dE dt}(E_{\nu}) = 
\int_0^5 dz f_{\rm ev}(z) 
\int dE_p \frac{dn}{dt}(E_{\nu}, E_p, z)
\end{equation}
where $f_{\rm ev}(z)$ and $dn/dt (E_{\nu},E_p,z)$ are 
the source-evolution factor and 
the number of cosmogenic neutrinos with energy $E_{\nu}$ at the Earth 
produced by protons with energy $E_p$ injected from a source 
at redshift $z$ per time at $z=0$. 
The normalization factor of the neutrino flux and the value of $\alpha$ 
are determined by fitting the observed UHECR spectra.

\section{Results} \label{results}

In this section, the results of our calculations 
of the fluxes of cosmogenic neutrinos are described. 
First of all, we compare our calculated spectrum to a spectrum estimated 
by Ref. \cite{stanev06} to check our simplified treatment 
of photopion production and to investigate the difference 
of the neutrino spectra predicted from 
2 different IR/UV background models in Section \ref{comparison}. 
In Section \ref{normalization}, 
the normalization factors of the neutrino flux and 
the spectral indices of UHECR injection at sources are determined 
for every source-evolution model and transition scenario 
so that the calculated spectra 
of UHE protons best reproduce the observed spectra. 
Then, we investigate the parameter dependence of the neutrino fluxes and 
discuss the capability to judge which transition scenario is favored 
in Section \ref{fluxes}. 
The detectability of cosmogenic neutrinos is also discussed.

\begin{figure}[t]
\begin{center}
\includegraphics[width=0.95\linewidth]{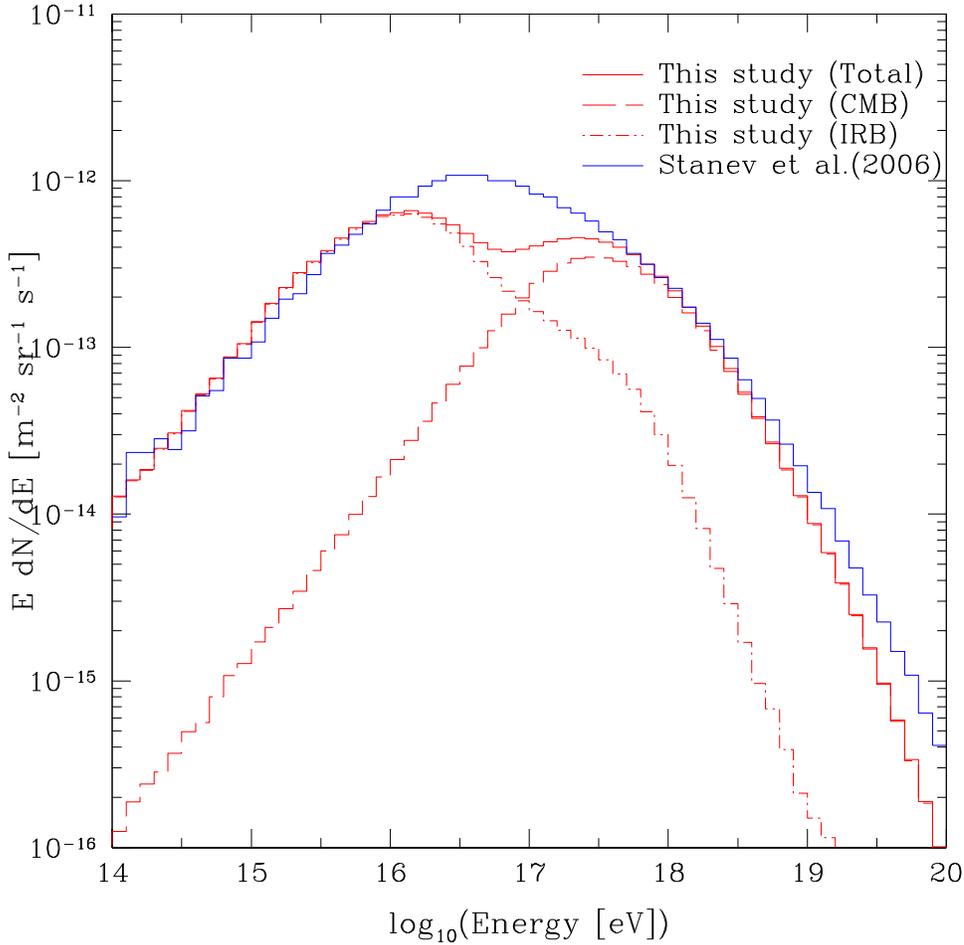}
\caption{Comparison of our $\nu_{\mu} + \bar{\nu_{\mu}}$ spectrum 
({\it red solid line}) to that calculated in Ref.\cite{stanev06} 
({\it blue solid line}). 
The same models and parameters are adopted for both calculations 
except for the treatment of photopion production and IR/UV background models.
Our result is divided into contributions 
by only the CMB ({\it red dashed line}) and 
that by the others ({\it red dot-dashed line}). 
The flux difference at $\sim 10^{17}$eV results from 
different IR/UV background models.}
\label{fig:comp}
\end{center}
\end{figure}

\subsection{Comparison with Another Estimation} \label{comparison}

Fig.\ref{fig:comp} shows a $\nu_{\mu} + \bar{\nu_{\mu}}$ 
spectrum calculated in this study ({\it red solid line}) 
and that in Ref.\cite{stanev06} ({\it blue solid line}). 
In addition, our spectrum is divided into two spectra: 
the contributions of the CMB photons ({\it red dashed line}) 
and the others ({\it red dot-dashed line}). 
Differences between the two calculations are the treatment 
of photopion production and IR/UV background models. 
Photopion production is treated by a simplified method explained 
in Section \ref{method} in this study, 
while fully simulated by using an event generator SOPHIA \cite{mucke00} 
in Ref. \cite{stanev06}. 
For IR/UV background photons, 
we adopt a model constructed by Ref. \cite{kneiske04}, 
whereas Ref. \cite{stanev06} used another model by Ref. \cite{stecker06}. 
The other parameters are the same ones for comparison: 
$E_{\rm max} = 10^{21.5}$eV, $E_{\rm min} = 10^{18}$eV,  
the proton injection spectrum of 
$dN/dE \propto E^{-2.5} \exp \left( -E/E_{\rm max} \right) 
\Theta(E - E_{\rm min})$, and 
a strong source evolution model defined in Ref.\cite{stanev06}. 
These fluxes are normalized by the same method as in Ref.\cite{stanev06}.

Our calculation well agrees with the estimation of Ref. \cite{stanev06}, 
but the neutrino fluxes at $\sim 10^{17}$ eV and above $10^{19.5}$ eV 
are about a factor of 2 smaller than those by Ref. \cite{stanev06}.

The discrepancy of the flux at $\sim 10^{17}$ eV is explained 
by the difference of adopted IR/UV background models as follows. 
At $z=0$, the number density of photons in the background model 
of Ref.\cite{kneiske04} is about twice as small as 
that of Ref.\cite{stecker06} at $\sim 0.2$ eV. 
Since photopion production dominantly occurs through the delta resonance, 
$s \sim 1.6~{\rm GeV}^2$, 
the energy of protons which dominantly interact 
with photons of 0.2 eV is $\sim 2 \times 10^{18}$ eV 
and the energy of produced neutrinos is $\sim 10^{17}$ eV. 
The mean free path of these protons for photopion production 
is $\sim 10^5$ Mpc at $z=0$ (see Fig. \ref{fig:mfp}), 
and therefore the Universe is transparent against these protons. 
Thus, the number of interactions is proportional to the photon number density, 
and then the IR/UV background radiation model by Ref.\cite{stecker06} 
leads to the neutrino flux twice as large as 
that in the background model used in this study at $\sim 10^{17}$ eV . 
Note that these neutrinos are not generated at high-redshift Universe 
because the energy-loss length of Bethe-Heitler pair-creation is 
much shorter than the interaction length 
of photopion production (see Fig.\ref{fig:mfp}). 
Hence, a discussion above at $z=0$ is sufficient. 
The shape of the spectrum of cosmogenic neutrinos 
gives us useful information on IR background photons.

The flux difference above $10^{19.5}$ eV originates 
from our simplified treatment of photopion production. 
Our treatment predicts a little lower flux than in Ref. \cite{stanev06} 
at the highest energy.

\begin{figure}[t]
\begin{center}
\includegraphics[clip,width=0.95\linewidth]{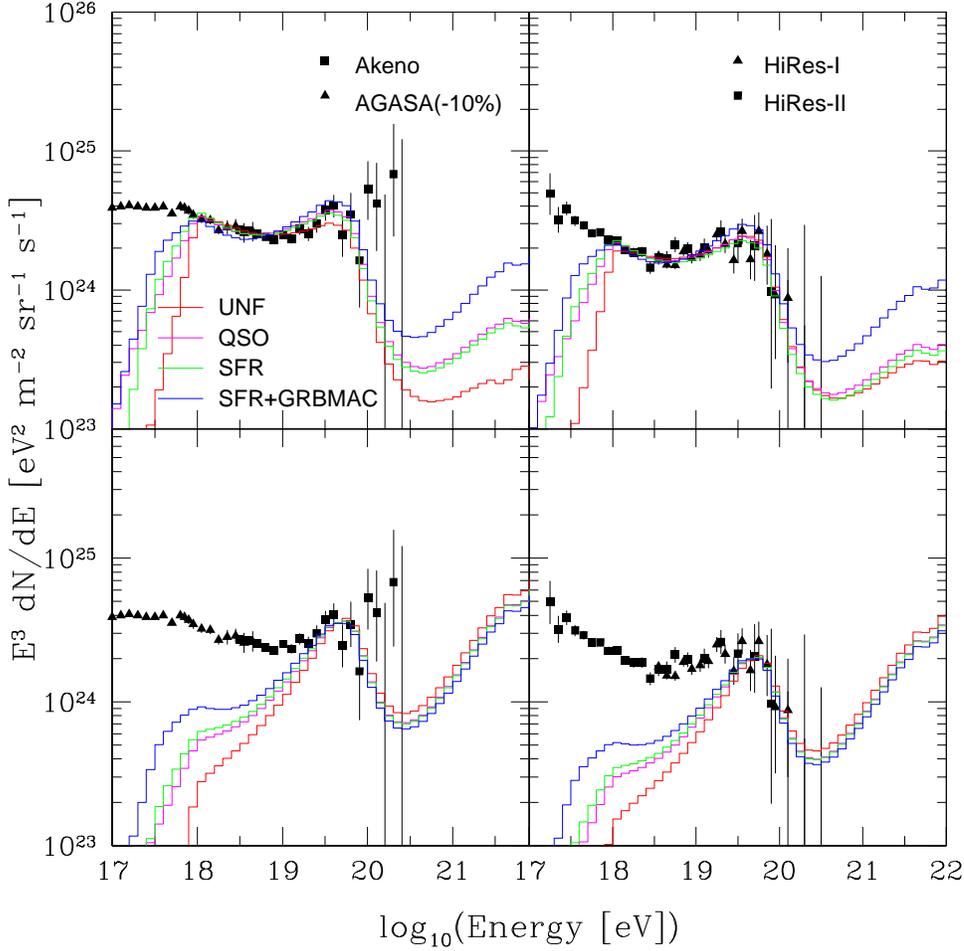}
\caption{Proton spectra fitted to the Akeno-AGASA data ({\it left two panels}) 
and HiRes data ({\it right two panels}). 
Note that a spectrum by the AGASA is shifted by 10\% 
to the lower energy to give a good agreement with the Akeno spectrum. 
$E_{\rm max}$ and $E_{\rm min}$ are set 
to be $10^{22}~{\rm eV}$ and $10^{18}$ eV, respectively. 
The upper two panels correspond to the proton-dip scenario. 
The spectral indices of proton sources are 2.7, 2.5, 2.5, and 2.3 
for UNF, QSO, SFR, and SFR+GRBMAC source-evolution models, respectively 
in the upper left panel. 
In the upper right panel, the spectral indices 
are 2.6, 2.5, 2.5, and 2.3, respectively. 
These are the best fit spectral indices of $\chi^2$ fittings 
with the observational spectra from $10^{18.0}$ to $10^{19.6}~{\rm eV}$. 
In the lower two panels, which are 
in the case of the ankle-transition scenario, 
the spectral indices are set to be 2.0.} 
\label{fig:proton}
\end{center}
\end{figure}

\subsection{Normalization of the Neutrino Flux} \label{normalization}

The calculated fluxes of cosmogenic neutrinos are normalized 
so as to give a good fit to observed UHECR spectra in this study. 
The spectral fit constrains the spectral indices 
of the proton injection spectra and the normalization factors 
of the neutrino fluxes through the normalization of UHECR fluxes 
at the same time. 
We adopt UHECR spectra observed 
by the Akeno \cite{nagano92,nagano95} and AGASA \cite{takeda03}, 
and by the HiRes-I and HiRes-II \cite{abbasi07} for the normalization. 
Note that the AGASA spectrum is shifted by 10\% 
to lower energies to give a good agreement 
with the Akeno spectrum \cite{nagano00}.

Fig.\ref{fig:proton} represents calculated proton spectra 
which are best fitted to the observed spectra 
for different source-evolution models listed in the figure. 
In the upper two panels, 
the calculated spectra are fitted using the chi-square method 
in the range from $10^{18.0}$ to $10^{19.6} {\rm eV}$, 
which correspond to the proton-dip scenario. 
As a result, 
spectral indices which best reproduce the Akeno-AGASA results 
are $\alpha =$ 2.7, 2.5, 2.5 and 2.3 for 
the UNF, QSO, SFR, and SFR+GRBMAC source-evolution models, respectively. 
Fitting to the HiRes spectra leads to $\alpha =$ 2.6, 2.5, 2.5, 
and 2.3, respectively. 
These indices are independent of $E_{\rm min}$ and $E_{\rm max}$ 
as long as $E_{\rm min} < 10^{18}$eV and $E_{\rm max} > 10^{20}$eV. 
In the lower two panels, 
we fix the spectral indices to 2.0 
and try to fit the calculated spectra to the observed ones 
in the energy range from $10^{19.5}$ to $10^{19.9}$ eV 
to represent the ankle-transition scenario. 
In both scenarios, shortfalls in the fluxes at lower energies 
are thought to be compensated by GCRs. 
Throughout this paper, 
we adopt the normalization based on the Akeno-AGASA spectrum.

The PAO also reported the energy spectra of UHECRs \cite{yamamoto07}. 
However, an energy spectrum derived from its ground-based detector 
covers only down to $10^{18.4}$eV. 
An energy spectrum constructed from a hybrid method covers above $10^{18}$eV, 
but it has a large statistical error due to the small number 
of detected events above $10^{19}$eV. 
Thus, we do not use the PAO spectra in this study.

\begin{figure}[t]
\begin{center}
\includegraphics[width=0.48\linewidth]{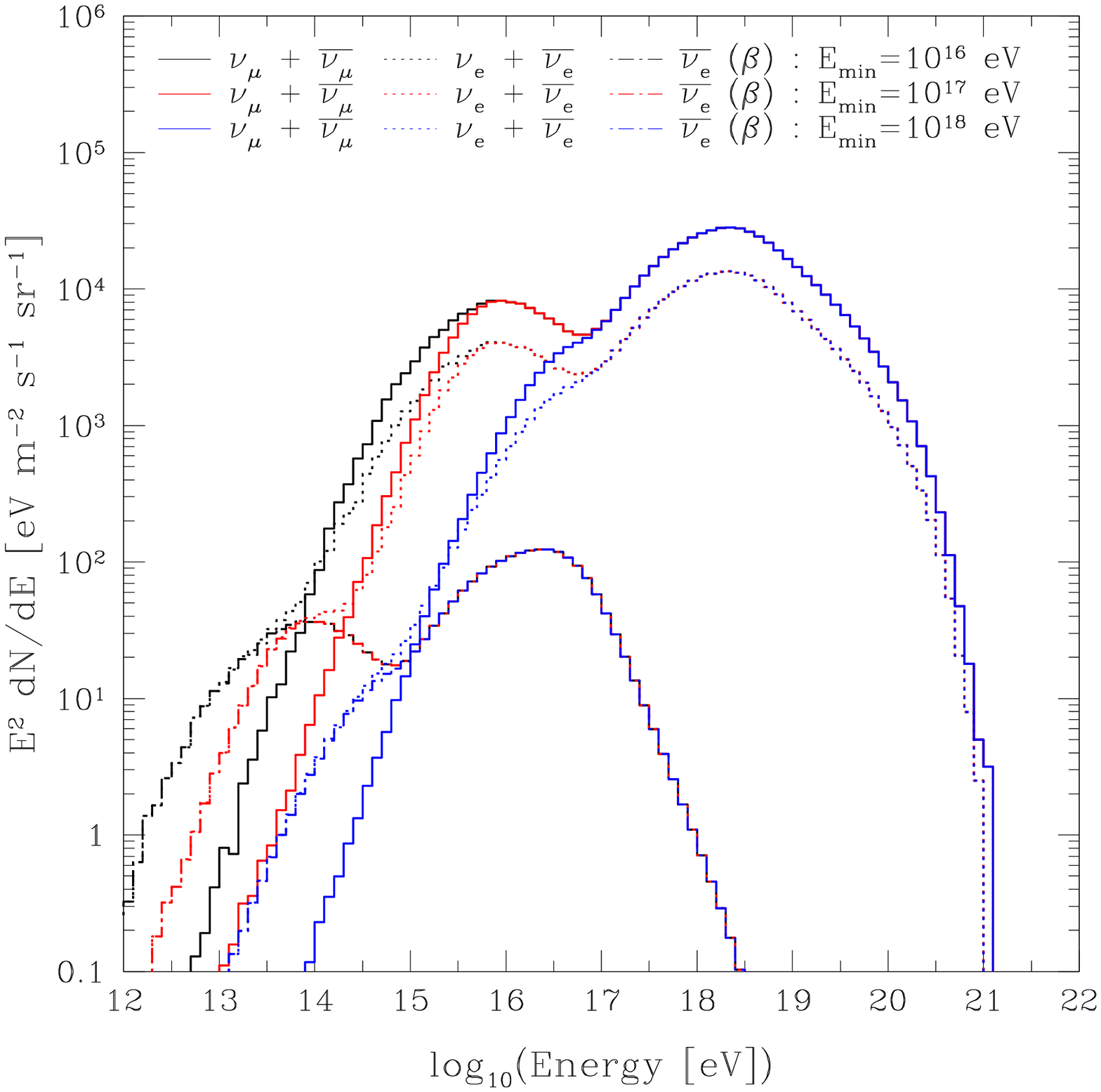} \hfill
\includegraphics[width=0.48\linewidth]{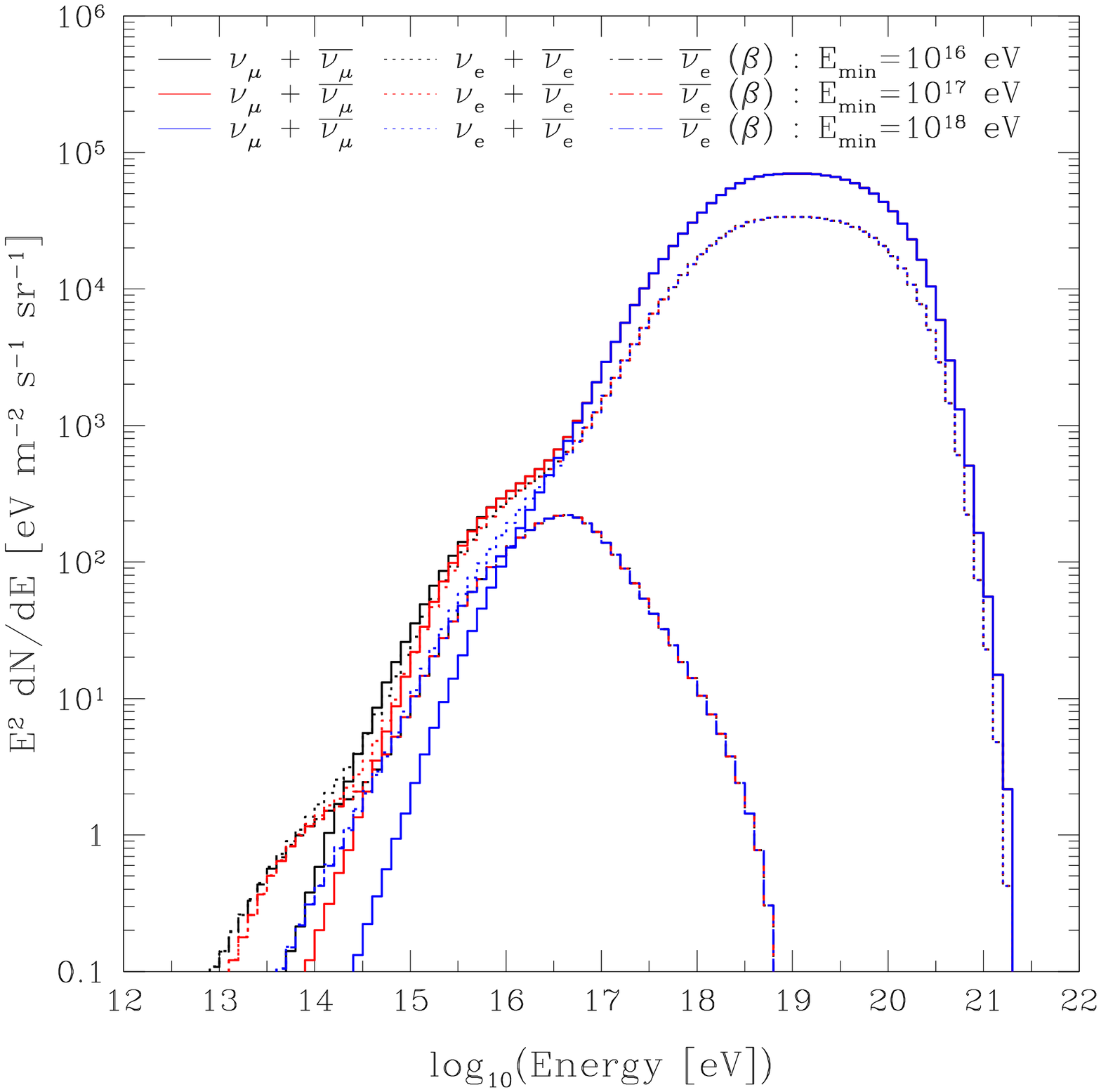}
\caption{Spectra of cosmogenic neutrinos 
with $E_{\rm min}=10^{16}$ ({\it black})
$10^{17}$ ({\it red}) and $10^{18}$eV ({\it blue}) 
in the proton-dip scenario ({\it left}) and the ankle-transition 
scenario ({\it right}). 
The solid lines, dotted lines, and dot-dashed lines 
show the spectra of $\nu_{\rm \mu} + \bar{\nu_{\rm \mu}}$, 
$\nu_e + \bar{\nu_e}$, and $\bar{\nu_e}$ from neutron beta decay, 
respectively. 
The cosmological evolution of UHECR sources and neutrino oscillation 
are not taken into account. 
All fluxes are normalized by the Akeno-AGASA spectrum.}
\label{fig:emin}
\end{center}
\end{figure}

\subsection{Neutrino Fluxes} \label{fluxes}

We start by investigating the dependence 
of the fluxes of cosmogenic neutrinos on $E_{\rm min}$. 
Fig.\ref{fig:emin} shows the calculated spectra of 
$\nu_{\rm \mu} + \bar{\nu_{\rm \mu}}$ ({\it solid lines}), 
$\nu_{e} + \bar{\nu_{e}}$ ({\it dotted lines}), 
$\bar{\nu_{e}}$ from neutron beta decay ({\it dot-dashed lines}) 
for $E_{\rm min} = 10^{16}$ ({\it black}), $10^{17}$ ({\it red}), 
and $10^{18}$eV ({\it blue}). 
These spectra are calculated based on the proton-dip 
({\it left}) and ankle-transition ({\it right}) scenarios. 
The cosmological evolution of UHECR sources and 
neutrino oscillation are not taken into account. 
The $\nu_{e} + \bar{\nu_{e}}$ fluxes are twice as low as 
the $\nu_{\rm \mu} + \bar{\nu_{\rm \mu}}$ fluxes 
in the energy range where neutron beta decay does not 
contribute to $\bar{\nu_e}$ fluxes 
because pion decay produces two muon neutrinos and one electron neutrino.

An intriguing feature is the spectral peaks of 
both $\nu_{\rm \mu} + \bar{\nu_{\rm \mu}}$ and 
$\nu_{e} + \bar{\nu_{e}}$ spectra at $\sim 10^{16}$ eV in the left panel. 
These peaks are generated by interactions 
between protons with $\sim 10^{17}$ eV and UV photons. 
Since there are many protons at lower energies 
due to a steep injection spectrum, 
UV background photons significantly contribute to the neutrino flux. 
On the other hand, 
we cannot see a spectral peak at $\sim 10^{16}$ in the right panel. 
The spectral peaks are a characteristic feature in the proton-dip scenario.

We should take neutrino oscillation into account 
for discussions of the detectability of cosmogenic neutrinos, 
as we can only observe the spectrum of neutrinos 
which suffers from neutrino oscillation during their propagation. 
Neutrinos generated from charged pions and successive muon decay 
have a flavor ratio of $\nu_e : \nu_{\mu} : \nu_{\tau} = 1:2:0$. 
Neutrino oscillation changes this ratio into $\sim 1:1:1$ \cite{learned95}. 
The flavor ratio approximately holds $1:1:1$ even at low energies 
unless anti-neutrinos from neutron $\beta$ decay are dominated.

\begin{figure}[t]
\begin{center}
\includegraphics[clip,width=0.95\linewidth]{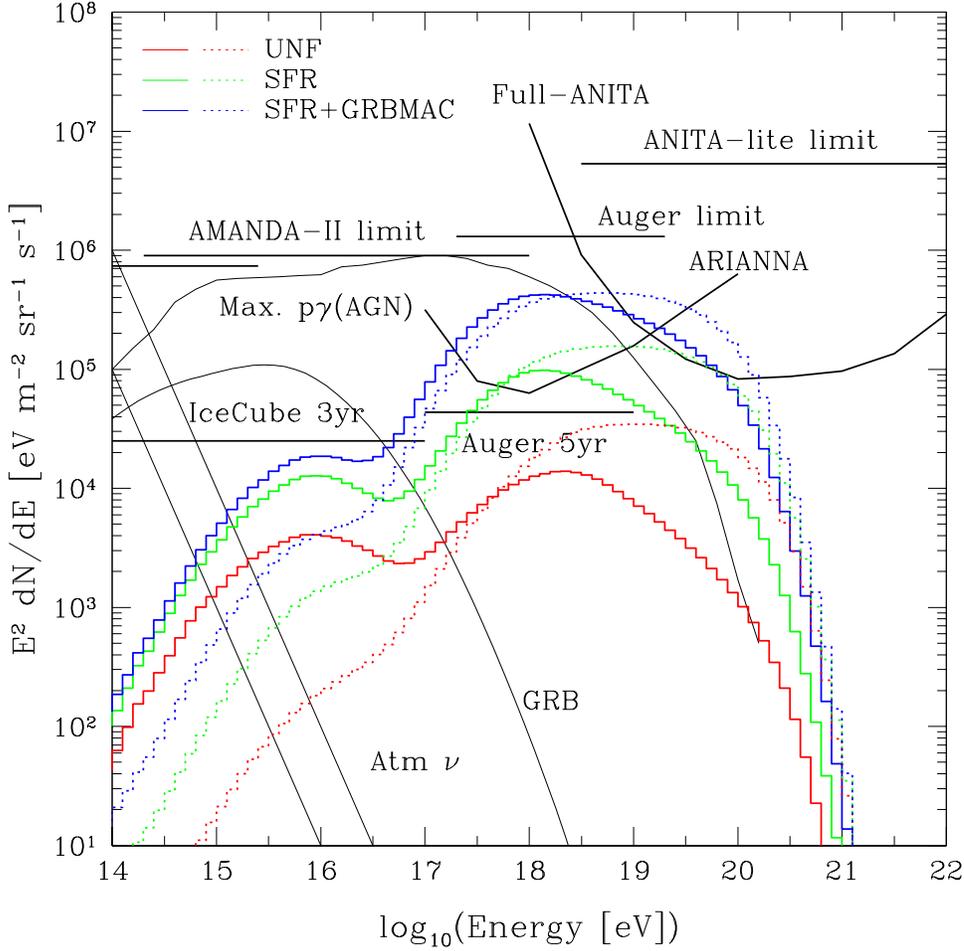}
\caption{Predicted spectra of cosmogenic neutrinos 
per flavor $( \nu_i + \bar{\nu}_i)$ 
in the proton-dip scenario ({\it solid lines}) 
and ankle-transition scenario ({\it dotted lines}). 
These fluxes are normalized by using the Akeno-AGASA spectrum. 
$E_{\rm min}$ and $E_{\rm max}$ are set to $10^{16}$ and $10^{22}$eV. 
The red lines, green lines, and blue lines are 
neutrino spectra for the UNF, SFR, 
and SFR+GRBMAC source-evolution models, respectively. 
The fluxes of the atmospheric neutrinos 
(Atm $\nu$ vertical/horizontal) \cite{lipari93} are represented. 
As upper limits of neutrino fluxes by several experiments, 
AMANDA-II limits \cite{amanda07a,amanda07b}, 
limit on tau neutrinos by the PAO \cite{pao07b}, 
ANITA-lite limit \cite{barwick06a} are shown. 
As estimated or projected sensitivities, 
sensitivity of 3 years observation by IceCube \cite{icecube_h}, 
sensitivity of 5 years observation by the PAO \cite{bertou02}, 
ARIANNA sensitivity \cite{barwick06b}, and
full-ANITA sensitivity \cite{barwick06a} are also shown. 
As diffuse neutrino spectra from energetic sources, 
a maximal neutrino flux from active galactic nuclei 
including neutrino oscillation \cite{mannheim01} 
and neutrino spectrum from GRBs calculated in Ref.\cite{murase06}, 
considering neutrino oscillation, 
with their parameters of $E_{\rm jet} = 1.24 \times 10^{51}~{\rm erg}$, $E_{\rm sh} = 10^{51}~{\rm erg}$, $\xi_{\rm B} = 1$, $\xi_{\rm acc} = 100$, $\Gamma = 10^{2.5}$, $r = 10^{13}-10^{14.5}~{\rm cm}$ and $l = r/\Gamma = 10^{10.5}~{\rm cm}$, which are used in Ref.\cite{achterberg07}, are shown.}
\label{fig:main}
\end{center}
\end{figure}

Fig.\ref{fig:main} shows the spectra of cosmogenic neutrinos 
per flavor, $\nu_i + \bar{\nu_i}$, taking neutrino oscillation into account. 
$E_{\rm min}$ and $E_{\rm max}$ are set to $10^{16}$ and $10^{22}$eV, 
respectively. 
Upper limits of neutrino fluxes determined by several experiments 
and the sensitivities of current/future neutrino detectors are also displayed. 
Theoretically estimated fluxes of diffuse neutrinos 
from AGNs \cite{mannheim01} and GRBs \cite{murase06} are also shown.

As mentioned above, 
the proton-dip scenario predicts a spectral peak at $\sim 10^{16}$ eV, 
while the peak does not appear for the ankle-transition scenario. 
The flux difference at $10^{16}$ eV between the two scenarios 
is about an order of magnitude. 
Therefore, the detection of this peak could be an evidence 
of the proton-dip scenario. 
The flux of the peak depends on source-evolution models. 
The SFR and SFR+GRBMAC source-evolution models result in 
three and four times higher neutrino fluxes 
than the UNF source-evolution model, 
respectively. 
In these strong source-evolution models, 
the predicted fluxes are comparable with the IceCube 3yr sensitivity.

The neutrino flux at $\sim 10^{20}$ eV could be also a clue 
to judge a better transition scenario. 
The ankle-transition scenario leads to the neutrino flux 
much higher than the proton-dip scenario at the highest energy 
because of a harder injection spectrum. 
In the SFR source-evolution model, 
the estimated flux reaches the full-ANITA sensitivity. 
Remember that our estimation of the flux of cosmogenic neutrinos 
above $10^{19.5}$eV might be underestimated by about a factor of 2 
as shown in Section \ref{comparison}. 
This works positively for the detection of the highest energy neutrinos. 
The full-ANITA is also expected to detect the highest energy neutrinos 
given the SFR+GRBMAC model is realistic enough, 
and it implies that GRBs are UHECR sources, though we should keep in mind that 
the models of the GRB rate history include large uncertainty
\footnote{A preliminary upper bound of the ANITA approaches the predicted flux 
of cosmogenic neutrinos with $E_{\rm max}=10^{22}$ eV 
in the ankle-transition scenario at $\sim 10^{20}$ eV 
(see {\it http://www.slac.stanford.edu/econf/C070730/}).}.

\begin{figure}[t]
\begin{center}
\includegraphics[clip,width=0.95\linewidth]{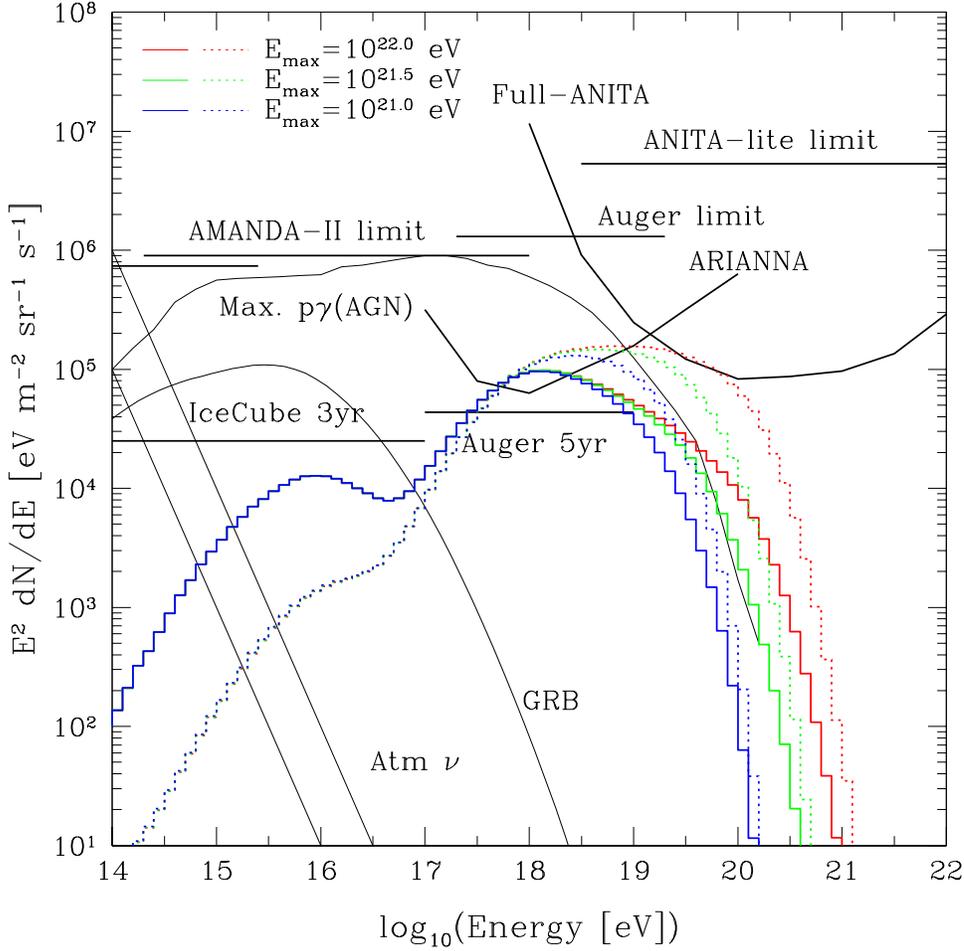}
\caption{The same as Fig.\ref{fig:main}, 
but spectra only calculated for the SFR source-evolution model 
in the cases of different $E_{\rm max}$, 
set to $10^{22}$ ({\it red lines}), $10^{21.5}$ ({\it green lines}), 
and $10^{21} {\rm eV}$ ({\it blues lines}). 
The solid lines are spectra based on the proton-dip scenario, 
and the dotted lines are those on the ankle-transition scenario.}
\label{fig:emax}
\end{center}
\end{figure}

The neutrino flux at the highest energy strongly depends on $E_{\rm max}$. 
Fig.\ref{fig:emax} shows the spectra of cosmogenic neutrinos 
calculated for $E_{\rm max} = 10^{22.0}$, $10^{21.5}$, and $10^{21.0}$eV 
in the SFR source-evolution model. 
All of $E_{\rm max}$ can reproduce the observed cosmic ray spectra sufficiently. 
We can see that the neutrino fluxes above $10^{19}$eV 
are sensitive to $E_{\rm max}$. 
As $E_{\rm max}$ becomes smaller, 
the neutrino flux at the highest energy is smaller and the flux difference 
between the two scenarios also becomes smaller. 
In order that the neutrino flux at $\sim 10^{20}$ eV can be a clue 
to distinguish between the two scenarios,  
sufficiently large $E_{\rm max}$ is required.

Whereas the neutrino fluxes at $\sim 10^{18}$ eV are independent 
of not only the transition scenarios (see Fig.\ref{fig:main}) 
but also $E_{\rm min}$ and $E_{\rm max}$ 
(see Figs.\ref{fig:emin} and \ref{fig:emax}), 
the neutrino fluxes are sensitive to source-evolution models. 
Thus, the neutrino flux at $\sim 10^{18}$ eV has information on 
the cosmological evolution of UHECR sources. 
The ARIANNA and PAO will detect cosmogenic neutrinos 
as long as the SFR or SFR+GRBMAC source-evolution scenarios are good ones. 
We can also obtain indirect knowledge on UHECR sources 
by comparing source-evolution models constrained by the neutrino observatories 
to the cosmological evolution models of various astrophysical objects.

The spectral features of cosmogenic neutrinos described above 
might be covered by the neutrino background from powerful objects 
like GRBs and AGNs, as shown in Fig.\ref{fig:main}. 
If the neutrino fluxes from these energetic objects are larger 
than the flux of cosmogenic neutrinos, 
it would be difficult to test transition scenarios by cosmogenic neutrinos.

A diffuse neutrino flux from GRBs estimated by Ref. \cite{murase06} 
is larger than 
the predicted fluxes of cosmogenic neutrinos up to $\sim 10^{17}$ eV 
and hides the spectral peaks of cosmogenic neutrinos at $\sim 10^{16}$ eV 
which is predicted in the proton-dip scenario. 
However, since neutrinos from GRBs are, in principle, distinguishable 
from cosmogenic neutrinos by time and spatial correlations 
between prompt $\gamma$-rays and neutrinos, 
the spectral peak can be detected.

For a diffuse neutrino flux from AGNs, 
the prediction of a maximal contribution model by Ref. \cite{mannheim01} 
is shown in Fig. \ref{fig:main}. 
The estimation covers not only the peak of the spectrum of cosmogenic neutrinos 
at $\sim 10^{16}$ eV for the proton-dip scenario 
but also the neutrino flux at $\sim 10^{18}$ eV. 
Unfortunately, 
spatial correlation between emitted neutrinos and photons 
is not expected for distant AGNs, 
because AGNs are not as bright as GRBs in general. 
However, we should notice that this flux is estimated as a maximum. 
The total flux of diffuse neutrinos starts to be restricted 
by AMANDA-II \cite{amanda07b} and will be constrained more precisely 
by observations in the near future by detectors such as IceCube. 
Whether the spectral features of cosmogenic neutrinos are not covered 
by AGN diffuse neutrinos and are observable will be determined by observations. 


\section{Discussion} \label{discussion}

In Section \ref{results}, 
we found that the spectrum of cosmogenic neutrinos can be an indicator 
to distinguish the two GCR/EGCR transition scenarios 
though its capability depends on several parameters on UHECR sources: 
$E_{\rm min}$, $E_{\rm max}$, and cosmological evolution models 
of UHECR sources. 
In this section, 
we discuss other uncertainties on the flux of cosmogenic neutrinos 
and their detectability.

The difference between the Akeno-AGASA and HiRes spectra 
brings the uncertainty of the neutrino flux 
through the normalization of the UHECR flux. 
The UHECR flux observed by the HiRes is twice as low as 
the Akeno-AGASA as shown in Fig.\ref{fig:proton} 
as long as the systematic errors of the determination of UHECR energies 
of both experiments are neglected. 
The flux difference is reflected to the flux of cosmogenic neutrinos directly, 
and therefore the difference of the neutrinos fluxes 
between the two normalizations is about a factor of 2.

A calibration of the energy-scale of UHECR observations could decrease 
the uncertainty of the neutrino flux. 
A dip calibration method in the proton-dip scenario 
\cite{berezinsky05,aloisio07} leads to a good agreement 
between the Akeno-AGASA and HiRes spectra 
assuming that the HiRes spectra are shifted to higher energy 
by a factor of 1.2 and the AGASA spectrum is shifted to lower energy 
by a factor of 0.9. 
Since the latter shift has already been performed 
for giving a good agreement with the Akeno spectrum, 
the normalization of the neutrino flux based on the shifted HiRes spectra 
results in the neutrino flux comparable with the Akeno-AGASA normalization 
if the proton-dip scenario is true.

Since we have not know which transition scenario is favored yet, 
the uncertainty on the neutrino flux which originates from 
the difference of observed UHECR fluxes is maximally a factor of 2.

The number density of IR/UV background photons also causes uncertainty 
on the flux of cosmogenic neutrinos. 
We already discussed the difference of the neutrino fluxes 
between different IR/UV background models of Refs. \cite{kneiske04,stecker06}. 
Since the number density of infrared photons 
in the model of Ref. \cite{stecker06} is twice as large as 
that of Ref. \cite{kneiske04} at low redshift, 
the former model predicts the neutrino flux twice as large as 
the latter model at $\sim 10^{17}$ eV. 
Here, we also compare the best-fit model, which is adopted in this study, 
with a low-infrared model in the same reference. 
At $z=0$, the number density of photons in the range of 0.005-0.5eV 
in the best-fit model is about twice as large as 
that in the low-infrared model. 
This energy range of photons corresponds to the neutrino energy 
of $4 \times 10^{16}$-$4 \times 10^{18}$eV. 
Thus, a neutrino flux predicted in the low-infrared model 
is twice as small as that in the best-fit model in this energy range. 
In fact, the CMB photons mainly contribute to the total neutrino flux 
above a few $\times 10^{17}$ eV. 
The low-infrared model predicts the neutrino flux twice as small as 
the best-fit model at $\sim 10^{16}$-$10^{17}$ eV. 
Adding the result of the discussion in Section \ref{comparison} to 
the discussion above, the uncertainty of the neutrino flux 
at $\sim 10^{16}$-$10^{17}$ eV is a factor of 4.

\begin{figure}[t]
\begin{center}
\includegraphics[clip,width=0.95\linewidth]{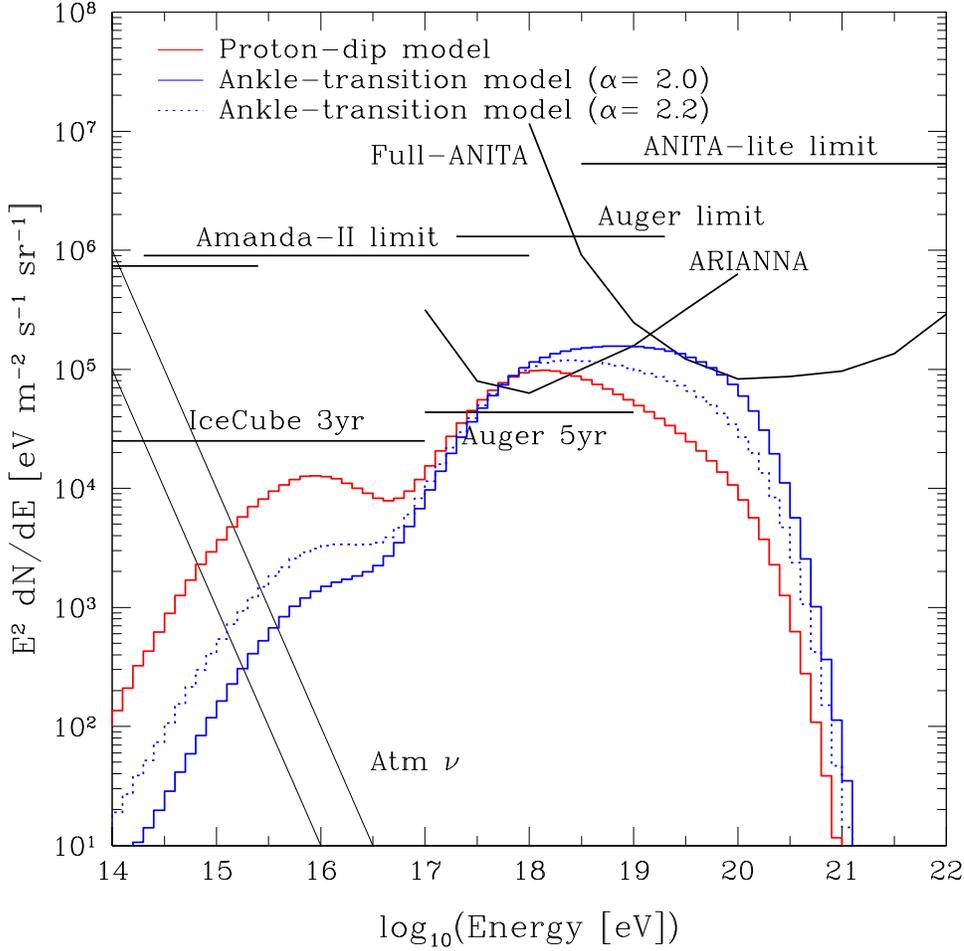}
\caption{Comparison between cosmogenic neutrino spectra 
with different spectral indices in the ankle-transition scenario. 
The blue solid line and blue dotted line are neutrino spectra 
with $\alpha = 2.0$ and $2.2$, respectively. 
The red line is neutrino spectrum in the proton-dip scenario for reference. 
These spectra are calculated in the SFR source-evolution model 
with $E_{\rm max}=10^{22}$eV and $E_{\rm min}=10^{16}$eV. 
Solid black lines are the same as those in Fig.\ref{fig:main}}
\label{fig:ankle}
\end{center}
\end{figure}

The difference of the spectral index of UHECR injection 
also slightly changes the neutrino flux. 
The spectral index has been fixed to 2.0 for the ankle-transition scenario 
in Section \ref{results}, 
but somewhat steeper injection spectrum is also allowed 
as mentioned in Section \ref{introduction}. 
We check the variation of the neutrino flux resulting from the change 
of the spectral index. 
Fig. \ref{fig:ankle} shows the comparisons of the neutrino fluxes 
predicted from two spectral indices in the ankle-transition scenario. 
For reference, a predicted spectrum in the proton-dip scenario is also shown. 
The flux difference between the two transition scenarios becomes smaller 
for a steeper spectrum in the ankle-transition scenario, 
which is about a factor of 3 at $\sim 10^{16}$ and $10^{20}$eV. 
In that case, 
the distinction between the two scenarios becomes little clear.

The composition of UHECRs could affect the flux of cosmogenic neutrinos. 
The composition is poorly known as mentioned in Section \ref{introduction}. 
Here, we consider only a mixed composition model adopted 
in Ref. \cite{allard06}. 
This model is an improved version of the ankle-transition scenario. 
Fig.9 of Ref.\cite{allard06} shows that nuclei heavier than protons contribute 
to the neutrino spectrum mainly below $10^{15} {\rm eV}$. 
Since the neutrinos below $10^{15} {\rm eV}$ are covered by atmospheric 
neutrinos as shown in Fig.\ref{fig:main} and \ref{fig:emax}, 
protons make a dominant contribution to the flux of cosmogenic neutrinos 
in an observable energy range in the ankle-transition scenario. 
Thus, the capability to distinguish the transition scenarios is 
not dramatically affected.

Finally, we revisit diffuse neutrino fluxes from AGNs and GRBs 
and discuss the detectability of cosmogenic neutrinos. 
In Fig.\ref{fig:main}, the diffuse neutrino flux from AGNs, 
maximally estimated in Ref.\cite{mannheim01}, is shown. 
This estimation assumes that GeV-TeV $\gamma$-rays from AGNs are of 
hadronic origin. 
This scenario requires high-energy hadrons which can generate neutrinos 
with energies comparable to cosmogenic neutrinos. 
Therefore, a large flux of high-energy neutrinos is predicted. 
On the other hand, the $\gamma$-rays can also be reproduced 
by leptonic origin (for review see \cite{aharonian04}). 
This scenario does not always predict UHE neutrinos 
because UHECR generation is not required. 
It is still controversial which model is the favorable one. 
Thus, whether the features of cosmogenic neutrino spectrum can be observed 
or not is related to the origin of high-energy $\gamma$-rays.

GRBs are also possible candidates of the sources of UHE neutrinos. 
In principle, high-energy neutrinos from GRBs can be distinguished 
from cosmogenic neutrinos 
since such neutrinos correlate with the prompt $\gamma$-rays of GRBs. 
However, recently, it has been suggested 
that neutrinos from low-luminosity GRBs can contribute 
to high-energy diffuse neutrino background \cite{murase06b}. 
Such neutrinos might be difficult to distinguish 
since the prompt emissions from such GRBs are difficult to observe. 
That neutrino flux can be comparable to the flux of cosmogenic neutrinos, 
$\sim 10^4~{\rm eV m}^{-2}{\rm sr}^{-1}{\rm s}^{-1}$ 
at $\sim 10^{16}~{\rm eV}$, 
but it strongly depends on the local rate of the low-luminosity GRBs, 
which has large uncertainties.

As discussed above, 
diffuse neutrinos from several powerful objects could prevent us 
from observing the spectral features of cosmogenic neutrinos. 
It is determined by future observations 
whether neutrinos from these objects can hide the spectral features.

\section{Conclusion} \label{conclusion}

We calculated the fluxes of cosmogenic neutrinos 
for several plausible parameter sets 
and discussed the possibility that the neutrino flux helps 
to judge which scenario of transition from GCRs to EGCRs is favored. 
We found that the spectrum of cosmogenic neutrinos has a characteristic peak 
at $\sim 10^{16}$ eV in the proton-dip scenario 
as long as extragalactic protons significantly contribute to 
the observed cosmic rays down to $10^{17}$ eV. 
The predicted flux is comparable with the sensitivity of IceCube 
when the SFR source-evolution model is considered. 
On the other hand, 
we also found that the neutrino flux at $\sim 10^{20}$ eV is 
much larger in the ankle-transition scenario than in the proton-dip scenario 
if the maximum energy of protons generated at sources is sufficiently high. 
If the SFR+GRBMAC source-evolution model is appropriate one, 
neutrinos with $\sim 10^{20}$ eV are expected to be detected by Full-ANITA 
though the flux is highly dependent on $E_{\rm max}$. 
These spectral features give us clues to judge which scenario is favored 
unless these are covered by the neutrino background 
from potential neutrino sources like AGNs and GRBs.

We also found that the neutrino flux at $\sim 10^{18}$ eV depends on 
only the cosmological evolution of UHECR sources. 
This indicates that the neutrino flux at this energy brings us 
information on the cosmological evolution of UHECR sources. 
The detection of this feature is feasible by ARIANNA and 
PAO if UHECR sources cosmologically evolve 
like star formation rate.

As discussed in this paper, 
the spectrum of cosmogenic neutrinos depends on many unknown parameters: 
$E_{\rm max}$, $E_{\rm min}$, source-evolution models, 
scenarios of the GCR/EGCR transition, the shape of the SED 
of IR/UV background photons. 
That is why cosmogenic neutrinos are the messengers 
of not only the nature of the EGCR sources 
but also cosmic background radiation. 
The future detection of cosmogenic neutrinos will provide us 
a lot of information on the Universe related to the highest energy phenomena.

\subsubsection*{Acknowledgements:} 
 
We thank Tanja Kneiske for publishing the tables 
of cosmological evolution of the infrared background from Ref.\cite{kneiske04}
at her website. 
We are grateful to Motohiko Nagano, Masahiro, Teshima, 
and Masaki Fukushima for providing Akeno data and useful comments. 
We are thankful to Shunsaku Horiuchi for helpful comments. 
The works of H.T. and K.M. are supported by Grants-in-Aid for JSPS Fellows. 
This work is partially supported by Grants-in-Aid for Scientific 
Research from the Ministry of Education, Culture, Sports, Science and 
Technology of Japan through No.19104006 (K.S. and S.N.), No.19047004,
and No.19740139 (S.N.), 
Grant-in-Aid for the 21st Century COE 
'Center for the Diversity and Universality in Physics' 
from the Ministry of Education, Culture, Sports, Science and Technology 
(MEXT) of Japan, and World Premier International Research Center 
Initiative (WPI Initiative), MEXT, Japan.


\begin{thebibliography}{}
\bibitem{stanev04} T. Stanev, High Energy Cosmic Rays, Springer-Varlag, 2004, pp.11
\bibitem{berezinsky05} V. Berezinsky, A. Gazizov, S. Grigorieva, Phys. Lett. B612 (2005) 147
\bibitem{aloisio07} R. Aloisio \etal, Astropart. Phys. 27 (2007) 76
\bibitem{abbasi05b} R.U. Abbasi, Astrophys. J. 622 (2005) 910
\bibitem{unger07} M. Unger \etal, arXiv:0706.1495
\bibitem{berezinsky69} V. Berezinsky, G. Zatsepin, Phys. Lett. B 28 (1969) 423
\bibitem{stecker73} F.W. Stecker, Astrophys. Space Sci. 20 (1973) 47
\bibitem{stecker79} F.W. Stecker, Astrophys. J. 228 (1979) 919
\bibitem{yoshida93} S. Yoshida, M. Teshima, Prog. Theor. Phys. 89 (1993) 833
\bibitem{engel01} R. Engel, D. Seckel, T. Stanev,  Phys. Rev. D 64 (2001) 093010
\bibitem{seckel05} D. Seckel, T. Stanev, Phys.~Rev.~Lett. 95 (2005) 141101
\bibitem{ave05} M. Ave \etal, Astropart. Phys. 23 (2005) 19
\bibitem{stanev06} M. de Marco \etal, Phys. Rev. D 73 (2006) 043003
\bibitem{allard06} D. Allard \etal, JCAP 09 (2006) 005
\bibitem{kneiske04} T.M. Kneiske \etal,  Astron. Astrophys. 413 (2004) 807
\bibitem{stecker06} F.W. Stecker, M.A. Malkan, S.T. Scully, Astrophys. J. 648 (2006) 774
\bibitem{stanev06b} T. Stanev, astro-ph/0611633
\bibitem{engel07} R. Engel \etal, arXiv:0706.1921
\bibitem{glushkov07} A.V. Glushkov \etal, JETP Lett. 87 (2008) 190
\bibitem{pao07a} The Pierre Auger Collaboration, Science 318 (2007) 938
\bibitem{pao07c} The Pierre Auger Collaboration, Astropart. Phys. 29 (2008) 188
\bibitem{berezinsky88} V.S. Berezinsky, S.I. Grigorieva, Astron. Astrophys. 199 (1988) 1
\bibitem{protheroe96} R.J. Protheroe, P.A.Johnson, Astropart. Phys. 4 (1996) 253
\bibitem{agostinelli03} S. Agostinelli \etal, 2003 Nucl. Instrum. Methods Phys. Res. A 506 (2003) 250, {\it http://wwwasd.web.cern.ch/wwwasd/geant4/geant4.html}
\bibitem{greisen66} K. Greisen, Phys. Rev. Lett. 16 (1966) 748
\bibitem{zatsepin66} G.T. Zatsepin, V.A. Kuz'min V A, JETP Lett. 4 (1966) 78
\bibitem{stanev00} T. Stanev \etal, Phys. Rev. D 62 (2000) 093005
\bibitem{chodorowski92} M.J. Chodorowski, A.A. Zdziarske, M. Sikora, Astrophys. J. 400 (1992) 181
\bibitem{murase06c} K. Murase S. Nagataki, Phys. Rev. Lett. 97 (2006) 051101
\bibitem{schadmand03} S. Schadmand, Eur. Phys. J. A 18 (2003) 405
\bibitem{asano06} K. Asano, S. Nagataki, Astrophys. J. Lett. 640 (2006) 9
\bibitem{waxman99} E. Waxman, J. Bahcall, Phys.Rev. D 59 (1999) 023002
\bibitem{waxman95} E. Waxman, Phys. Rev. Lett. 75 (1995) 386
\bibitem{guetta07} D. Guetta, T. Piran, JCAP 07 (2007) 003
\bibitem{jimenez06} R. Jimenez, Z. Haiman, Nature 440 (2006) 501
\bibitem{stanek06} K.Z. Stanek \etal, Acta Astron. 56 (2006) 333
\bibitem{fruchter06} A.S. Fruchter \etal, Nature 441 (2006) 463
\bibitem{yuksel06} H. Y$\ddot{\rm u}$ksel, M.D. Kistler, Phys. Rev. D 75 (2007) 083004
\bibitem{yoon06} S.C. Yoon, N. Langer, C. Norman, Astron. Astrophysics 460 (2006) 199
\bibitem{berezinsky02} V. Berezinsky \etal, astro-ph/0204357
\bibitem{berezinsky06} V. Berezinsky \etal, Phys. Rev. D 74 (2006) 043005
\bibitem{mucke00} A. Mucke \etal, Comput.~Phys.~Commun. 124 (2000) 290
\bibitem{nagano92} M. Nagano \etal, J.~Phys.~G.: Nucl.~Part.~Phys. 18 (1992) 423
\bibitem{nagano95} M. Nagano, 1995 private communication
\bibitem{takeda03} M. Takeda \etal, Astropart. Phys. 19 (2003) 447
\bibitem{abbasi07} R.U. Abbasi \etal, Phys. Rev. Lett. 100 (2008) 101101
\bibitem{nagano00} M. Nagano, A.A. Watson, Rev. Mod. Phys. 72 (2000) 689
\bibitem{yamamoto07} T. Yamamoto \etal, arXiv:0707.2638
\bibitem{learned95} J.G. Learned, S. Pakvasa, Astropart. Phys. 3 (1995) 267
\bibitem{lipari93} P. Lipari, Astropart. Phys. 1 (1993) 195
\bibitem{amanda07a} A. Achterberg \etal, Phys. Rev. D 76 (2007) 042008
\bibitem{amanda07b} M. Ackermann \etal, Astrophys. J. 675 (2008) 1014
\bibitem{pao07b} The Pierre Auger Collaboration, Phys. Rev. Lett. 100 (2008) 211101
\bibitem{barwick06a} S.W. Barwick \etal, Phys. Rev. Lett. 96 (2006) 171101
\bibitem{icecube_h} IceCube {\it http://www.icecube.wisc.edu/}
\bibitem{bertou02} X. Bertou \etal, Astropart. Phys. 17 (2006) 183
\bibitem{barwick06b} S.W. Barwick, astro-ph/0610631
\bibitem{mannheim01} K. Mannheim, R.J. Protheroe, J.P. Rachen, Phys. Rev. D 63 (2000) 023003
\bibitem{murase06} K. Murase, S. Nagataki, Phys. Rev. D 73 (2006) 063002
\bibitem{achterberg07} A. Achterberg \etal, Astrophys. J. 664 (2007) 397
\bibitem{aharonian04} F.A. Aharonian, Very High Energy Cosmic Gamma Radiation, World Scientific, 2004, pp.402
\bibitem{murase06b} K. Murase \etal, Astrophys. J. Lett. 651 (2006) 5
\end{thebibliography}
\end{document}